\documentclass{article}
\usepackage{spconf,amsmath,graphicx,hyperref}
\usepackage{bm} 
\usepackage{adjustbox}
\usepackage{multirow}
\usepackage{booktabs}

\title{LipSody: Lip-to-Speech Synthesis with Enhanced Prosody Consistency}
\name{Jaejun Lee$^{\star}$\thanks{This paper has been accepted to ICASSP 2026.} \qquad Yoori Oh$^{\star}$ \qquad Kyogu Lee$^{\star \dagger \ddagger}$}
  
\address{$^{\star}$ Music and Audio Research Group (MARG),\\
Department of Intelligence and Information, Seoul National University\\
  $^{\dagger}$ Interdisciplinary Program in Artificial Intelligence, Seoul National University\\
  $^{\ddagger}$ Artificial Intelligence Institute, Seoul National University, Republic of Korea}
      
\begin{document}
\ninept
\maketitle
\begin{abstract}
Lip-to-speech synthesis aims to generate speech audio directly from silent facial video by reconstructing linguistic content from lip movements, providing valuable applications in situations where audio signals are unavailable or degraded. While recent diffusion-based models such as LipVoicer have demonstrated impressive performance in reconstructing linguistic content, they often lack prosodic consistency.
In this work, we propose LipSody, a lip-to-speech framework enhanced for prosody consistency. LipSody introduces a prosody-guiding strategy that leverages three complementary cues: speaker identity extracted from facial images, linguistic content derived from lip movements, and emotional context inferred from face video. Experimental results demonstrate that LipSody substantially improves prosody-related metrics—including global and local pitch deviations, energy consistency, and speaker similarity—compared to prior approaches.
\end{abstract}
\section{Introduction} \label{sec:intro}
Lip-to-speech synthesis aims to reconstruct or generate speech audio from silent video input, with most research focusing on recovering the original speech content by analyzing the visible articulatory movements of the speaker’s lips. This cross-modal task holds particular value in scenarios where the speech signal is absent or degraded, such as silent video footage, hearing-impaired communication aids, or as a complementary modality in multi-modal speech synthesis.

Benefiting from advances in deep learning, lip-to-speech synthesis has gained momentum with the development of end-to-end frameworks that directly map facial video frames to acoustic features. Most existing systems, however, have primarily focused on reconstructing linguistic content, with progress typically measured by intelligibility metrics such as word error rate (WER). Yet, given the rich and diverse visual cues present in facial video, it is equally important to generate speech that conveys the speaker’s plausible vocal characteristics. Beyond simply synchronizing lip movements, achieving natural and personalized speech synthesis requires realistic prosody and speaker identity that align with the speaker’s facial appearance and emotional expression.

Motivated by these observations, we present LipSody, a novel lip-to-speech framework that enhances the consistency of speaker-dependent prosody. Unlike previous works that focus primarily on intelligibility, LipSody leverages visual input not only for articulatory content but also to infer prosodic cues and speaker-specific voice characteristics.
Building upon the state-of-the-art LipVoicer~\cite{yemini2023lipvoicer}, our model explicitly estimates prosodic features—namely, pitch and energy—by conditioning on three complementary embeddings derived exclusively from facial video: lip movement-based linguistic content, face image-based speaker identity, and face video-based emotional expression.
Through this design, LipSody produces speech that is not only intelligible but also expressive and personalized, exhibiting improved prosodic consistency and speaker resemblance from purely visual input.

We summarize our contributions as follows:
\begin{itemize}
\item We present LipSody, a diffusion-based lip-to-speech framework that significantly improves prosody consistency while maintaining state-of-the-art intelligibility.
\item We introduce a novel visual-only prosody estimation method, which leverages three complementary cues: lip movement-based linguistic content, face image-based speaker identity, and face video-based emotional expression.
\item We conduct extensive evaluations using a diverse set of prosody-related metrics, demonstrating the effectiveness of our approach and uncovering a compelling correlation between prosody consistency and speech intelligibility.
\end{itemize}
The demo is available on, \url{https://jaejunL.github.io/LipSody_Demo/}.

\section{Related work} \label{sec:related}

\begin{figure*}[t!]
\centerline{\includegraphics[width=1.9\columnwidth]{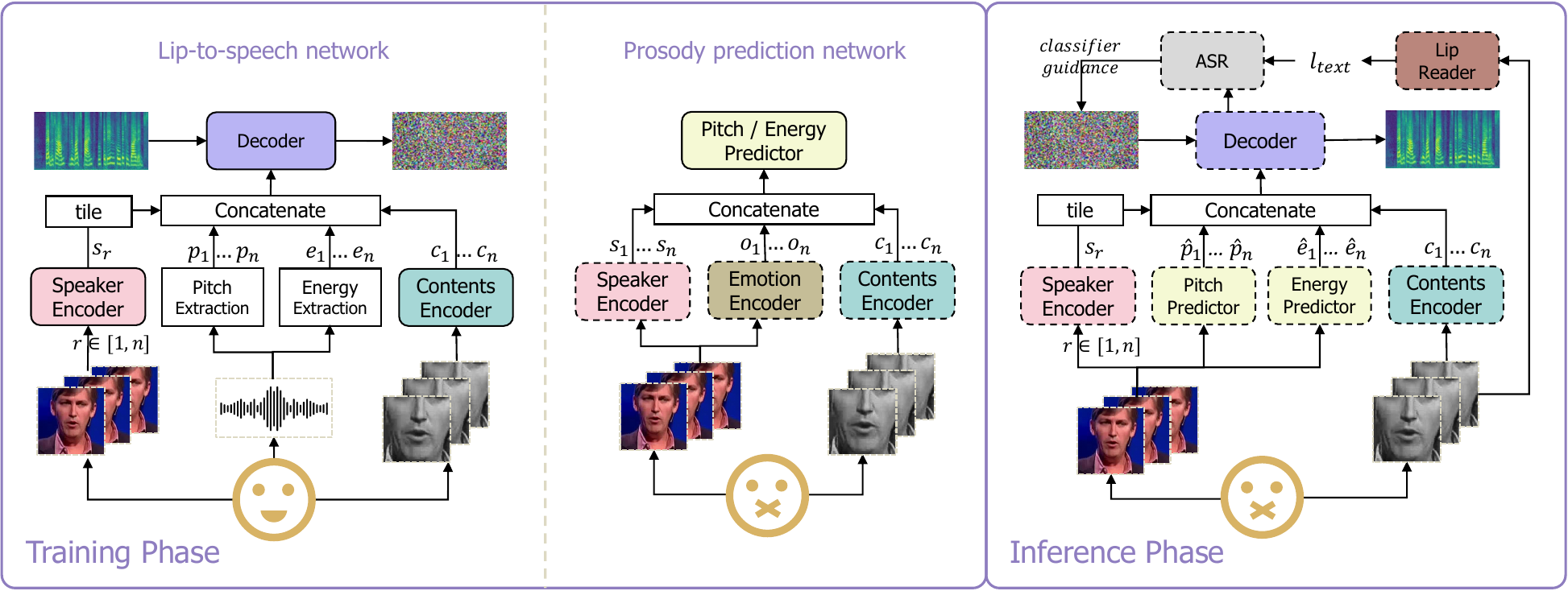}}
\caption{Overview of LipSody, the proposed diffusion-based lip-to-speech framework with enhanced pitch consistency. During training, ground-truth pitch and energy values are used to provide prosody-related supervision to the lip-to-speech network. For inference, an independent network is trained to predict these prosody features from lip movement-based linguistic content, face image-based speaker identity, and face video-based emotional expression.}
\label{fig:figure1}
\end{figure*}

\subsection{Lip-to-speech}
Lip-to-speech synthesis targets reconstructing speech audio from silent video frames. Early works trained CNN-RNN architectures to predict speech features directly from lip motion sequences~\cite{afouras2018deep}, primarily evaluating performance through word error rate (WER).

Recent efforts have aimed to improve the intelligibility and naturalness of generated speech through techniques such as GAN-based refinement
~\cite{hegde2022lip},
self-supervised unit representations~\cite{hsu2023revise, choi2023intelligible}, and speech decomposition methods~\cite{choi2025v2sflow}.
Notably, recent work such as LipVoicer~\cite{yemini2023lipvoicer} has shown strong results in speech intelligibility using diffusion-based speech generation~\cite{kong2020diffwave}. Their method leverages both classifier guidance (CG)~\cite{dhariwal2021diffusion} and classifier-free guidance (CFG)~\cite{ho2022classifier} to enhance conditional generation quality. Specifically, the model uses a pretrained lip-reading model~\cite{ma2021towards} to generate pseudo ground-truth text descriptions, and a pretrained automatic speech recognition (ASR) model~\cite{burchi2023audio} as the guiding classifier.

However, most lip-to-speech models still struggle to capture fine-grained prosodic features such as pitch and energy variation, resulting in less natural and personalized speech synthesis.

\section{Methods} \label{sec:methods}
We introduce LipSody, a lip-to-speech framework specifically designed to enhance prosody consistency, aiming to generate speech with pitch and energy patterns that are more faithfully aligned with the speaker’s visual expressions.

\subsection{Diffusion-based lip-to-speech network}
Our model is built upon the state-of-the-art diffusion-based lip-to-speech framework LipVoicer~\cite{yemini2023lipvoicer}, which has demonstrated significant improvements in intelligibility, notably reducing word error rate (WER).
LipVoicer generates a mel-spectrogram using a conditional Denoising Diffusion Probabilistic Model (DDPM)~\cite{ho2020denoising}, where conditional generation is guided by CFG~\cite{ho2022classifier} . The conditioning signal is derived solely from video input and consists of two components:
a full-face image~($\mathcal{S}$) used to extract speaker identity features ($\boldsymbol{s}$), and a lip-centered cropped video sequence~($\mathcal{C}$) used to extract linguistic content features ($\boldsymbol{c}$). For $\boldsymbol{s}$, a single frame is randomly selected from among the $n$ frames, whereas for $\boldsymbol{c}$, all $n$ frames are utilized.
During inference, to further enhance the intelligibility of the generated speech, LipVoicer employs CG~\cite{dhariwal2021diffusion}. Specifically, it uses a pretrained ASR model to compute the guidance signal: $\nabla_{\mathbf{x_t}} \log p(l \mid \mathbf{x}_t)$, where $\mathbf{x}_t$ is the intermediate mel-spectrogram at diffusion step $t$, and $l$ is the predicted text label from a pretrained lip-reading model. Finally, the generated mel-spectrogram is transformed into the waveform using a pretrained vocoder.

\subsection{Enhancing prosody consistency with LipSody}
Inspired by LipVoicer, we propose LipSody — a prosody-consistency enhanced diffusion-based lip-to-speech framework. Fig.~\ref{fig:figure1} illustrates the overall training and inference pipelines.
Similar to LipVoicer, LipSody is trained using a DDPM-based architecture with CFG.
However, beyond utilizing $\boldsymbol{s}$ and $\boldsymbol{c}$, we explicitly model prosodic characteristics by injecting frame-wise pitch ($\boldsymbol{p}$) and energy ($\boldsymbol{e}$) values, extracted from paired ground-truth speech during training, serving as oracle supervision signals.

To obtain the mel-spectrogram, it is common practice to normalize the original waveform $y$ by dividing it by its maximum absolute value within each clip. However, we hypothesize that each speaker possesses a unique energy distribution, and clip-wise normalization removes these speaker-specific characteristics.
Since the goal of LipSody is to enhance prosody consistency, we instead perform speaker-wise normalization.
Formally, the normalized waveform for the $j$-th clip of speaker $i$ is defined as $y_{\mathbf{norm}}^{ij} = y^{ij} / \max(\forall y^i)$, where $y^{ij}$ denotes the original waveform signal of the $j$-th clip belonging to speaker $i$, and the maximum is taken over all clips for that speaker.

During inference, we retain the CG strategy with lip-reader predicted text to preserve the strong intelligibility of LipVoicer.
Combining this with our proposed prosody modeling, the inferred noise $\boldsymbol{\epsilon}$ at each diffusion step is computed as:
\begin{equation}
\begin{aligned}
\boldsymbol{\epsilon} = (1 + w_1) \,
\boldsymbol{\epsilon}_\theta(\mathbf{x}_t, \boldsymbol{s}, \boldsymbol{\hat{p}}, \boldsymbol{\hat{e}}, \boldsymbol{c}) - w_1 \, \boldsymbol{\epsilon}_\theta(\mathbf{x}_t)\\
-w_2\,\sqrt{1-\bar{\alpha}_t}\nabla_{\mathbf{x_t}} \log p(l \mid \mathbf{x}_t),
\end{aligned}
\label{eq:lipsody}
\end{equation}
where $\boldsymbol{\hat{p}}$ and $\boldsymbol{\hat{e}}$ denote the predicted pitch and energy features, respectively, while $w_1$ and $w_2$ control the strengths of CFG and CG.
The therm $\bar{\alpha}=\prod_{s=1}^t\alpha_s$ is the cumulative product of the noise schedule coefficients, where $a_t=1-\beta_t$ and $\beta_t$ is the variance schedule.
Following LipVoicer, we also adopt gradient normalization when applying Eq.~\ref{eq:lipsody}.

\subsection{Prosody prediction} \label{subsec:pe}
The oracle $\boldsymbol{p}$ and $\boldsymbol{e}$ values are unavailable during inference. Therefore, we independently train prosody prediction networks to estimate each value from video-derived features.
Rather than relying solely on face image-based speaker identity~($\boldsymbol{s}$) and lip movement-based linguistic content ($\boldsymbol{c}$), we further enhance prosody estimation by incorporating rich emotional cues captured in the video modality. Specifically, we utilize a pretrained Emotion Encoder, which produces an emotion embedding $\boldsymbol{o}$.
During training of the lip-to-speech network, only a randomly selected single frame is used to extract $\boldsymbol{s}$, meaning that $\boldsymbol{s}$ carries relatively coarse-grained speaker information.
We hypothesize that the broad, time-dependent emotion embedding captures fine-grained information that can further refine frame-wise prosody prediction.
To estimate prosodic features, we concatenate $\boldsymbol{s}$, $\boldsymbol{o}$, and $\boldsymbol{c}$ and feed the combined representation into the prosody prediction networks. Both the Speaker Encoder and Content Encoder within the prosody prediction network are pretrained from the base LipSody lip-to-speech model, while only the pitch and energy predictors (Fig.~\ref{fig:figure1}) remain trainable during this stage. Each predictor is optimized independently using a mean squared error (MSE) loss.

\subsection{Model architecture}
\noindent\textbf{Contents Encoder}: 3D convolutional network similar to~\cite{ma2021towards}\\
\textbf{Speaker Encoder}: ResNet-18~\cite{he2016deep} with the final two fully connected layers discarded.\\
\textbf{Decoder}: A DiffWave-based decoder consisting of stacked dilated residual blocks with gated activations, conditioned on mel-spectrogram and prosodic features~\cite{kong2020diffwave}.\\
\textbf{Pitch Extraction}: Fast Context-base Pitch Estimator (FCPE)~\footnote{ https://github.com/CNChTu/FCPE}
with the normalization similar to SoVITS~\footnote{https://github.com/svc-develop-team/so-vits-svc}
\\
\textbf{Energy Extraction}: Average across the frequency axis of a log-mel spectrogram.\\
\textbf{Prosody Predictor}: Self-attention layers followed by feed-forward layers.\\
\textbf{Emotion Encoder}: An EmoCLIP-based encoder leveraging contrastive text--image pretraining~\cite{foteinopoulou2024emoclip}.\\
\textbf{ASR}: Audio-visual conformer network~\cite{burchi2023audio}, though only audio input is utilized.\\
\textbf{Lip Reader}: Conformer-based lip-reading model architecture~\cite{ma2023auto}\\
\textbf{Vocoder}: HiFi-GAN architecture~\cite{kong2020hifi}.

\begin{table}[t!]
\caption{Evaluation results of conventional metrics}
\centering
\begin{adjustbox}{width=0.5\textwidth}
\begin{tabular}{cccccc}
\hline\hline
& $\mathrm{WER}\downarrow$ & $\mathrm{STOI}\uparrow$
& $\mathrm{DNSMOS}\uparrow$
& $\mathrm{LSE}\text{-}D\downarrow$ & $\mathrm{LSE}\text{-}C\uparrow$\\
\hline
$\mathrm{LipVoicer_{official}}$ & {0.219} & {0.93} & {3.19} & {8.166} & {6.308} \\
$\mathrm{LipVoicer_{recon}}$ & {0.221} & {0.93} & {3.17} & {8.516} & {6.400} \\ 
$\mathrm{LipSody}$ (ours) & {0.225} & {0.92} & {3.18} & {8.435} & {6.404} \\ 
\hline\hline
\end{tabular}
\end{adjustbox}
\label{tab:result1}
\vspace{-2mm}
\end{table}

\section{Experiments}
\subsection{Datasets}
We used LRS3 dataset~\cite{afouras2018lrs3}, which comprises 5,502 TED and TEDx videos totaling over 430 hours of speech. Each video is preprocessed to include a cropped facial region at a resolution of 224×224 pixels, with video sampled at 25 frames per second and audio recorded as single-channel 16kHz waveforms.
For training, we follow the predefined \textit{pretrain} and \textit{train} splits, and use the \textit{test} split for evaluation, adopting the unseen speaker protocol used in LipVoicer~\cite{yemini2023lipvoicer}.
For mouth-region cropping, we follow the approach in~\cite{ma2022visual}.
A 96×96 mouth-centered bounding box is extracted as the region of interest. The video patches are converted to grayscale and normalized using the training set’s global mean and variance.

\subsection{Comparison systems}
As our primary objective is to enhance prosody while maintaining high intelligibility, we exclude recent works that report relatively degraded WER performance (33.9\% in \cite{hsu2023revise}, 29.8\% in \cite{choi2023intelligible}, and 28.5\% in \cite{choi2025v2sflow}). Instead, we focus our comparison on the state-of-the-art framework LipVoicer~\cite{yemini2023lipvoicer} (WER 21.9\%), evaluating our method against two variants of it.

The first is the official model reported in the original paper, specifically the version using a HiFi-GAN~\cite{kong2020hifi} vocoder, denoted as $\mathrm{LipVoicer_{official}}$.
The second is our reimplementation using the publicly available codebase with default training settings, denoted as $\mathrm{LipVoicer_{recon}}$.
Our proposed model, which incorporates explicit prosody modeling, is referred to as $\mathrm{LipSody}$.

\subsection{Implementation details}
For preprocessing and the base lip-to-speech model, we follow the official LipVoicer implementation~\cite{yemini2023lipvoicer}, including the diffusion setup with $T=400$ denoising steps and a variance schedule of $\beta_1=0.0001$ to $\beta_T=0.02$.
The Prosody Predictor and Emotion Encoder are built upon public implementations from SoVITS and EmoCLIP~\cite{foteinopoulou2024emoclip}, respectively.
Training runs for 100K steps with a batch size of 256, following the original protocol.
Regarding CFG and CG, the weights $w_1$ and $w_2$ are fixed to 2 and 1.5, respectively, following the original configuration of~\cite{yemini2023lipvoicer}.

\begin{table}[t!]
\caption{Evaluation results of prosody-related metrics}
\centering
\begin{adjustbox}{width=0.5\textwidth}
\begin{tabular}{cccccc}
\hline\hline
& $\mathrm{GF}_0\downarrow$ & $\mathrm{LF}_0\downarrow$ & $\mathrm{EC}\downarrow$ & $\mathrm{Resem}\uparrow$ & $\mathrm{Resem}_{tv}\uparrow$\\
\hline
$\mathrm{LipVoicer_{recon}}$ & {28.87} & {45.26} & {1.2667} & {0.5974} & {0.6578} \\ 
$\mathrm{LipSody}$ (ours) & {$\textbf{25.15}$} & {$\textbf{41.06}$} & {$\textbf{0.9141}$} & {$\textbf{0.6040}$} & {$\textbf{0.6618}$} \\ 
\hline\hline
\end{tabular}
\end{adjustbox}
\label{tab:result2}
\vspace{-2mm}
\end{table}


\subsection{Metrics}
We evaluate our models using both objective and subjective metrics. The objective evaluation consists of two types:
(1) conventional metrics commonly used in lip-to-speech literature to assess intelligibility and quality, and
(2) prosody-related metrics that we introduce to evaluate the model’s ability to capture prosodic information, such as pitch and energy consistency. For subjective evaluation, we conducted a naturalness test and an ABX test using Amazon Mechanical Turk (MTurk). In the ABX test, raters judged which synthesized speech sample had prosody more similar to the ground truth.

\subsubsection{Objective metrics (Conventional)} \label{subsec:convmetric}
\textbf{WER}: Calculated using an ASR model~\cite{ma2023auto} to assess the intelligibility of generated speech.\\
\textbf{STOI}: A non-intrusive intelligibility metric estimated using STOI-Net~\cite{zezario2020stoi}.\\
\textbf{DNSMOS}: A non-intrusive neural MOS predictor for perceptual speech quality~\cite{reddy2022dnsmos}.\\
\textbf{LSE}: Lip-sync error measured by SyncNet~\cite{chung2017out}, with two components: $\mathrm{LSE\text{-}D}$ (lip–audio distance) and $\mathrm{LSE\text{-}C}$ (sync confidence).

\subsubsection{Objective metrics (Prosody-related)} \label{subsec:prosodymetric}
\textbf{GF$_0$} (Global $F_0$): L1 loss between the synthesized average pitch and the speaker’s global $F_0$ (avg. over all clips)~\cite{lee2024hear}, assessing speaker-specific pitch statistics.\\
\textbf{LF$_0$} (Local $F_0$): L1 loss between frame-wise pitch contours of synthesized and ground-truth speech~\cite{lee2025speaking}, evaluating fine-grained, content-aligned pitch variation.\\
\textbf{EC} (Energy Consistency): L2 loss between frame-wise energy values from log-mel spectrograms, averaged over frequency.\\
\textbf{Resem}: Cosine similarity between speaker embeddings of synthesized and reference audio, using Resemblyzer~\footnote{https://github.com/resemble-ai/Resemblyzer}.\\
\textbf{Resem$_{tv}$}: Time-varying Resem with windowed cosine similarity (2s window, 1s hop), assessing dynamic speaker identity.

\subsubsection{Subjective metrics} \label{subsec:subject}
\textbf{Naturalness}: Evaluating the perceptual quality of synthesized audio using a 5-point MOS scale (from completely unnatural to completely natural). Each task was assigned to 100 subjects with a reward of 0.5 USD.\\
\textbf{ABX test}: Measuring subjective preference between two models, where participants were asked to judge which synthesized sample exhibited prosody more consistent with the paired ground-truth audio. Participants were instructed to focus only on prosodic aspects such as pitch, energy, and voice style, while ignoring pronunciation errors or audio quality. Each task was distributed to 100 subjects, with a reward of 2 USD.

\begin{table}[t!]
\caption{Evaluation results of subjective metrics}
\centering
\begin{adjustbox}{width=0.4\textwidth}
\begin{tabular}{cccccc}
\hline\hline
& $\mathrm{Naturalness}\uparrow$ & $\mathrm{ABX\space test(\%)}\uparrow$\\
\hline
$\mathrm{GT speech}$ & {4.02} & {-}\\
\hline
$\mathrm{LipVoicer_{recon}}$ & {3.36} & {45.78}\\ 
$\mathrm{LipSody}$ (ours) & {3.47} & {$\textbf{54.22}$}\\ 
\hline\hline
\end{tabular}
\end{adjustbox}
\label{tab:resultsub}
\vspace{-2mm}
\end{table}

\section{Results}
\subsection{Evaluation with objective conventional metrics}
The results of the conventional metric evaluation are presented in Table~\ref{tab:result1}.
First, a comparison between $\mathrm{LipVoicer_{official}}$ and $\mathrm{LipVoicer_{recon}}$ shows that the reconstructed model achieves performance comparable to the official implementation across all metrics. Although $\mathrm{LipVoicer_{recon}}$ exhibits slightly poor $\mathrm{WER}$ and $\mathrm{LSE}$, the differences are marginal.
When comparing $\mathrm{LipSody}$ with $\mathrm{LipVoicer_{recon}}$, we observe no statistically significant difference across any of the conventional metrics ($p>0.05$ in paired t-tests).
These results demonstrate that LipSody retains high intelligibility and perceptual quality, validating that our prosody-enhancement strategy does not compromise baseline performance.

\subsection{Evaluation with objective prosody-related metrics} \label{subsec:result2}
Table~\ref{tab:result2} presents the results for prosody-related metrics, which directly evaluate the main objective of our proposed method. In all five prosody metrics, LipSody consistently outperforms LipVoicer ($p<0.01$ in paired t-tests). The improvements in $\mathrm{GF}_0$ and $\mathrm{LF}_0$ indicate that LipSody better captures speaker-specific vocal traits and generates more natural, content-aligned pitch contours—purely from visual input. The gains in the $\mathrm{EC}$ metric further support this finding, indicating improved modeling of frame-wise speech dynamics associated with prosody. Finally, the results of $\mathrm{Resem}$ and $\mathrm{Resem}_{tv}$ demonstrate that LipSody not only preserves speaker identity in a coarse, clip-level sense but also captures its time-varying fluctuations, reflecting more faithful and expressive voice synthesis.

Note that $\mathrm{EC}$ is computed non-parametrically from the mel-spectrogram, whereas $\mathrm{GF}_0$ and $\mathrm{LF}_0$ are estimated with FCPE, a parametric pitch extractor. To ensure that our findings are not tied to this specific estimator, we additionally computed $\mathrm{GF}_0$ and $\mathrm{LF}_0$ using several alternative parametric extractors~\cite{wei2023rmvpe, kim2018crepe}. Across all schemes, LipSody consistently and significantly outperformed LipVoicer. Due to space constraints, detailed results are omitted.

\subsection{Evaluation with subjective metrics}
Table~\ref{tab:resultsub} presents the results of the naturalness and ABX test. For naturalness, LipSody achieved slightly higher scores than LipVoicer, but the difference was not statistically significant ($p>0.05$ in paired t-tests). This outcome is expected, as our primary contribution lies in enhancing prosody while maintaining intelligibility; thus, achieving comparable naturalness to the benchmark is reasonable.
In the ABX preference test, which measures prosody consistency with the ground truth, LipSody significantly outperformed LipVoicer. We further tested whether the ABX accuracy exceeded the chance level of 0.5 using a one-sample t-test, and found that our model achieved a statistically significant improvement over chance ($p < 0.05$).
These subjective evaluation results reinforce the effectiveness of our method in enhancing prosody consistency while preserving intelligibility.

\subsection{Performance According to Prosody Information}
Table~\ref{tab:result3} compares the performance of three variants of the LipSody framework. The first variant corresponds to the full $\mathrm{LipSody}$ configuration used in Table~\ref{tab:result1} and Table~\ref{tab:result2}. The second variant, denoted with `$w/o$ emotion $\boldsymbol{o}$', removes the Emotion Encoder from the prosody prediction network (Fig.~\ref{fig:figure1}). In this setting, the prosody predictor receives only the speaker embedding $\boldsymbol{s}$ and content embedding $\boldsymbol{c}$, excluding the emotion embedding $\boldsymbol{o}$.
Compared to the full LipSody model, it shows a statistically significant degradation in prosody-related metrics—including $\mathrm{GF_0}$, $\mathrm{LF_0}$, and $\mathrm{EC}$ ($p<0.01$ in paired t-tests). This suggests that incorporating emotion-related features derived from facial image sequences contributes meaningfully to accurate prosody estimation.
The third variant, denoted with `$w$ oracle~$\boldsymbol{p}$,$\boldsymbol{e}$', uses the ground-truth pitch $\boldsymbol{p}$ and energy $\boldsymbol{e}$ during inference. As expected, this configuration yields substantial improvements across prosody-related metrics, as it serves as an upper-bound reference where the true prosodic features are available. Interestingly, this setting also shows a notable improvement in WER. This result implies that more accurate prosody estimation not only enhances prosody consistency but also contributes to improved speech intelligibility.

\begin{table}[t!]
\caption{Results according to prosody information settings}
\centering
\begin{adjustbox}{width=0.42\textwidth}
\begin{tabular}{ccccc}
\hline\hline
& {$\mathrm{WER}\downarrow$} & {$\mathrm{GF}_0\downarrow$} & {$\mathrm{LF}_0\downarrow$} & {$\mathrm{EC}\downarrow$} \\
\hline
$\mathrm{LipVoicer_{recon}}$ & {0.221} & {28.87} & {45.26} & {1.2667} \\ 
$\mathrm{LipSody}$ (ours) & {0.225} & {25.15} & {41.06} & {0.9141} \\ 
$w/o$ emotion $\boldsymbol{o}$ & {0.220} & {27.11} & {41.26} & {0.9402} \\ 
$w$ oracle $\boldsymbol{p}$,$\boldsymbol{e}$ & {\textbf{0.210}} & {\textbf{16.98}} & {\textbf{31.11}} & {\textbf{0.4620}} \\ 
\hline\hline
\end{tabular}
\end{adjustbox}
\label{tab:result3}
\vspace{-2mm}
\end{table}

\section{Conclusion}
We introduced LipSody, a novel lip-to-speech synthesis framework that enhances prosody consistency by integrating visual-driven pitch and energy modeling into a diffusion-based generation pipeline. Unlike prior approaches that mainly emphasize intelligibility, LipSody explicitly leverages visual cues—speaker identity from facial images, linguistic content from lip movements, and emotional context from face video—for prosody modeling.

Our method achieves significant improvements in prosody-related metrics while maintaining strong intelligibility. Furthermore, we found that enhanced prosody modeling also contributes to intelligibility, underscoring the close interplay between expressive delivery and content clarity.

\vfill\pagebreak

\section{acknowledgement}
This research was supported by Basic Science Research Program through the National Research Foundation of Korea(NRF) funded by the Ministry of Education [No. RS-2025-25398143, 50\%], Institute of Information \& communications Technology Planning \& Evaluation (IITP) grant funded by the Korea government(MSIT) [No. RS2022-II220641, 45\%], and [No. RS-2021-II211343, 5\%]

\bibliographystyle{IEEEbib}
\bibliography{refs}

\end{document}